\documentclass[aip, jap, preprint]{revtex4-1}

\usepackage{graphicx}
\usepackage{epstopdf}
\usepackage{amsmath}
\usepackage{amssymb}
\usepackage{ulem}

\begin{document}
\title{Disorder regimes and equivalence of disorder types in artificial spin ice}

\author{Zoe~Budrikis}
\email{zoe.budrikis@gmail.com}
\affiliation{School of Physics, The University of Western Australia, 35 Stirling Hwy, Crawley 6009, Australia}
\affiliation{Istituto dei Sistemi Complessi CNR, Via Madonna del Piano 10, 50019 Sesto Fiorentino, Italy}
\affiliation{SUPA School of Physics and Astronomy, University of Glasgow, Glasgow G12 8QQ, United Kingdom}
\author{Paolo~Politi}
\affiliation{Istituto dei Sistemi Complessi CNR, Via Madonna del Piano 10, 50019 Sesto Fiorentino, Italy}
\affiliation{INFN Sezione di Firenze, via G. Sansone 1, 50019 Sesto Fiorentino, Italy}
\author{R.L.~Stamps}
\affiliation{SUPA School of Physics and Astronomy, University of Glasgow, Glasgow G12 8QQ, United Kingdom}

\maketitle

\section*{Abstract}
The field-induced dynamics of artificial spin ice are determined in part by interactions between magnetic islands, and the switching characteristics of each island. Disorder in either of these affects the response to applied fields. Numerical simulations are used
to show that disorder effects are determined primarily by the strength of disorder relative to inter-island interactions, rather than by the type of disorder. 
Weak and strong disorder regimes exist and can be defined in a quantitative way.

\newpage

Artificial spin ice~\cite{Wang:2006, Qi:2008} is a two-dimensional frustrated system, consisting of arrays of single-domain sub-micron islands of magnetic material, which are coupled by stray magnetic fields. 
It was conceived as an experimental model of three dimensional spin ices \cite{1997Harris} such as Dy$_2$Ti$_2$O$_7$ and Ho$_2$Ti$_2$O$_7$, with the advantage that its microstates can be imaged directly due to the relatively large island volume.
A consequence of this large volume is that artificial spin ice is typically athermal, and dynamics must be induced by an external magnetic field, which acts to ``tilt'' the complex energy landscape and allow transitions between configurational states.

Disorder in the system affects field-driven dynamics via two mechanisms. Disorder in island interactions (energetic disorder) alters the energies of configurational states. On the other hand, disorder in island magnetization switching characteristics (switching disorder) changes the barriers between states, altering the pathways the system takes through its space of configurational states~\cite{Budrikis:2011networks}.
There are many possible sources of quenched disorder. In this paper we study four disorder types and show that their effects are similar. We also distinguish two regimes of disorder strength, namely the weak disorder regime and the strong disorder regime.

Previous experimental studies of disorder have shown that the hysteresis of artificial spin ice reveals the importance of disorder \cite{Kohli:2011}. The switching characteristics of connected \cite{Mellado:2010, Ladak:2010} and disconnected \cite{Mengotti:2010} kagome ice arrays have been used to estimate the strength of disorder in those systems. In simulation studies of an analog to artificial spin ice based on vortices in patterned superconductors, disorder is shown to nucleate grain boundaries between different ground state orderings \cite{Libal:2009}.

The four disorder types studied here are disorder in island positions, disorder in island orientations, a random perturbation to pairwise interactions and a random perturbation to island switching fields. 
Pairwise energy disorder is a simplified `representation' of energetic disorder -- in particular, it does not lead to correlations in island pair energies, unlike positional and orientational disorder. 
In all cases, disorder is 
assumed to have a Gaussian distribution. For example, positional disorder is implemented by perturbing the $x$ and $y$ coordinates of each island $i$ by $\delta_{x,y}^{(i)}$, where the $\delta$ come from a Gaussian distribution with mean $0$ and standard deviation $\sigma$.

\begin{figure}[p]
  \centering
  \includegraphics{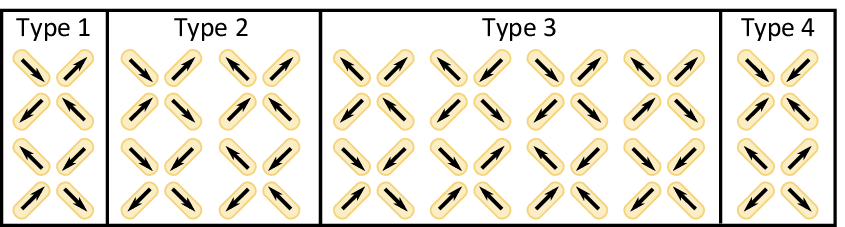}
  \caption{The sixteen vertex configurations of square artificial spin ice can be classified into four types, based on energy.}
  \label{vertex_types}
\end{figure}

We limit our considerations to square artificial spin ice \cite{Wang:2006}. In square geometry, four islands meet at each vertex of the array, as shown in Fig.~\ref{vertex_types}. The interactions of the four islands are not all equivalent: four of the six pairwise interactions are stronger than the other two, due to the difference in distances and orientations between islands. 
As a result, there are four energetically-degenerate types of vertex configuration, and the ground state of the system is a tiling of Type 1 vertices. 
The population fraction of Type 1 vertices, $n_1$,  is a good
measure of the energy of the system. 

In the absence of disorder, the four vertex types have distinct, single-valued energies $E_1<E_2<E_3<E_4$. However, when energetic disorder is present the energy of a vertex depends on the particular disorder realization. Considering many disorder realizations leads to a distribution of energies for each vertex type. The width of the distribution enables us to characterize the strength of disorder.

We make this characterization in a model system in which the island magnetizations are Ising point dipoles, interacting so that the dipolar energy of island $i$ is
\begin{equation}
\begin{split}
E_{\mathrm{dip}}^{(i)} &= -\vec h_{\mathrm{dip}}^{(i)}\cdot \vec{M}_i \\
	&= \frac{1}{4 \pi \mu_0} \sum_{j\ne i} \biggl(
	\frac{\vec{M_i} \cdot \vec{M_j}}{r_{ij}^3}
	- 3 \frac{(\vec{M_i} \cdot \vec{r_{ij}}) (\vec{M_i} \cdot \vec{r_{ij}})}{r_{ij}^5}
	\biggr)
\end{split}
\end{equation}
with the island magnetization $M$ and the nearest-neighbor distance both set to unity, so that the nearest-neighbor coupling has strength $1.5$, in units of $1/(4\pi\mu_0)$. In a perfect system, this gives $E_1=-6+\sqrt{2}$, $E_2=-\sqrt{2}$, $E_3=0$, $E_4=6+\sqrt{2}$.

\begin{figure}[p]
  \centering
  \includegraphics{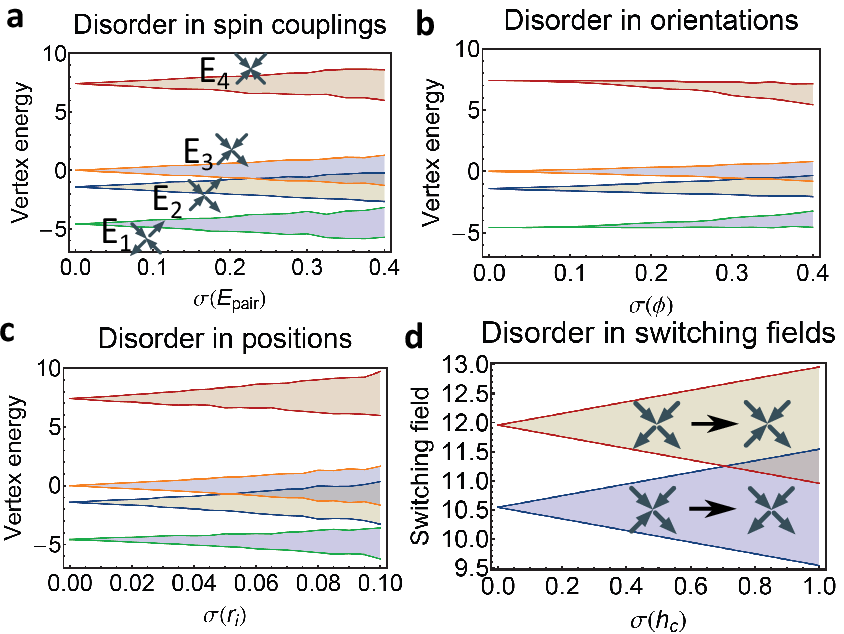}
  \caption{The mean energy plus/minus one standard deviation of a (top-to-bottom) Type 4, 3, 2, 1 vertex, subject to (a) pairwise energy, (b) orientational and (c) positional disorder. The point where the Type 2 and Type 3 bands cross is taken to be the transition between the weak and strong disorder regimes. The distributions are calculated over 200 independent disorder realisations. (d) The mean plus/minus one standard deviation of the external field required to convert a Type 2 to a Type 3 vertex (upper band) and the reverse (lower band). The insets show the two processes.}
  \label{disorder_strength}
\end{figure}

Shown in Fig.~\ref{disorder_strength}(a--c) are energy bands for each vertex type, defined by the mean energy plus/minus one standard deviation, plotted against disorder strength $\sigma$ for the three energetic disorder types. 
The bands become wider as the disorder strength increases, and eventually overlap occurs. We take the weak disorder regime to be that where there is no overlap, and the threshold of the strong disorder regime being the point where two bands first cross (in practice, the first crossing is always of the Type 2 and Type 3 bands, since they are closest initially). The transition occurs at $\sigma(E_{\mathrm{pair}})\approx 0.225$ 
for pairwise energy disorder, $\sigma(\phi)\approx0.3$ for island orientational disorder and $\sigma(r_i)\approx0.05$ for island positional disorder. (Note that $r_i=r_{x,y}$ refers to each component of an island's position.)

The fourth type of disorder we discuss has a different nature, because it
is disorder in island magnetization switching characteristics. We model the switching of the Ising moment of island $i$ as occuring when
\begin{equation}
\label{switching}
\vec h_{\mathrm{tot}}^{(i)}\cdot \hat m_i < -h_c,
\end{equation}
where $\vec h_{\mathrm{tot}}^{(i)}$ is the sum of the external field $\vec h$ and the dipolar field $\vec{h}_{\mathrm{dip}}^{(i)}$ and $\hat m_i$ is the direction of the island's magnetic moment.

Unlike energetic disorder, switching disorder leaves the energies of configurational states unchanged and cannot be characterized in terms of vertex energies. However, an analogous measure can be constructed by considering the applied field required for two single spin flip processes. 
When all switching fields are equal, the applied field required to convert a Type 2 vertex into a Type 3 vertex is $h_c + 1/\sqrt{2}$, while the field required for the reverse process is $h_c - 1/\sqrt{2}$. These processes are shown as insets to Fig.~\ref{disorder_strength}(d). 
We take the boundary between weak and strong disorder to be the meeting point of the plus/minus one standard deviation bands for the fields required for the two processes. The standard deviation of the switching fields, $\sigma(h_c)$, is equal to the standard deviation in the applied field required, so the crossing point can be easily determined to be $\sigma=1/\sqrt{2}\approx0.7$, as seen in Fig.~\ref{disorder_strength}(d). Note that this measure of disorder strength is relative to the inter-island coupling, not the mean switching field, since changing $h_c$ does not affect the disorder strength at which the bands meet. Thus, both energetic and switching disorder are measured relative to the same quantity and can be compared meaningfully. 

\begin{figure}[p]
  \centering
  \includegraphics{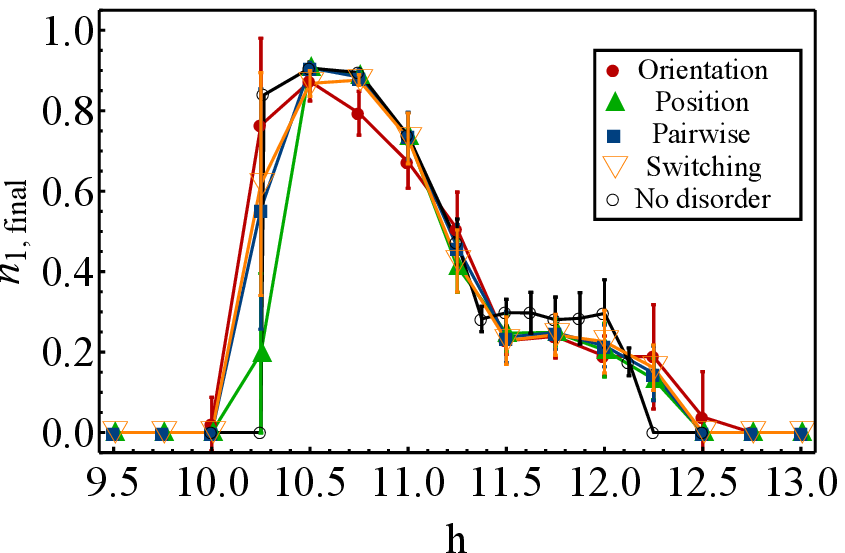}
  \caption{Final Type 1 population vs applied field $h$ for a perfect $20\times 20$ system (black open circles) and four different types of weak disorder: island position disorder with standard deviation in each component of the positions of $0.01$ (green filled triangles); island orientation disorder with standard deviation $0.06$ (red filled circles); island pair energies perturbed with standard deviation $0.045$ (blue filled squares); and switching field disorder with standard deviation $0.14$ (orange empty triangles). In all cases, the disorder strength $\sigma=0.2\sigma^*$, where $\sigma^*$ is the transition from weak to strong disorder.}
  \label{n1_vs_h_rotating_weak_dis}
\end{figure}

\begin{figure}[p]
  \centering
  \includegraphics{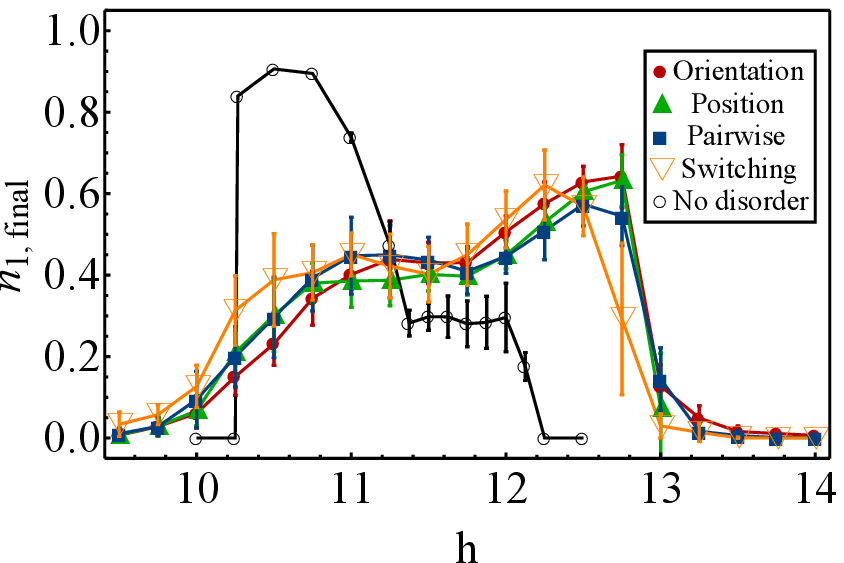}
  \caption{Final Type 1 population vs applied field $h$ for a perfect $20\times 20$ system with open edges (black open circles) and four different types of strong disorder: island position disorder with standard deviation in each component of the positions of $0.06$ (green filled triangles); island orientation disorder with standard deviation $0.36$ (red filled circles); island pair energies perturbed with standard deviation $0.27$ (blue filled squares); and switching field disorder with standard deviation $0.84$ (orange empty triangles). In all cases, the disorder strength $\sigma=1.2\sigma^*$, where $\sigma^*$ is the transition from weak to strong disorder.}
  \label{n1_vs_h_rotating_strong_dis}
\end{figure}

To see that the strength of disorder is more important than its origin, we compare in Figures \ref{n1_vs_h_rotating_weak_dis} and \ref{n1_vs_h_rotating_strong_dis} the final Type 1 population of a $20\times20$ island square ice, attained under a rotating applied field for both a perfect system and systems subject to different types of weak disorder (Fig.~\ref{n1_vs_h_rotating_weak_dis}) and strong disorder (Fig.~\ref{n1_vs_h_rotating_strong_dis}). In the simulations, the field rotates with constant amplitude and angular step $d\theta=0.01$ radians for 10 cycles, long enough to obtain a steady state. At each field application, the system relaxes by flipping single spins according to criterion~\eqref{switching} until no further flips are possible.

We have previously shown~\cite{Budrikis:2010} that, in the absence of disorder, dynamics that create Type 1 vertices can only occur for a narrow range of field amplitudes $\Delta h \simeq 2$ and dynamics always start at array edges. Smaller fields are unable to flip spins, and larger fields force the magnetization to track the field. Between these limits, there are two field regimes. In the low field regime, $10.25\lesssim h \lesssim 11.25$, dynamics proceed in a regular process of spin flips that ``invade'' from the array edges. In the high field regime, $11.25\lesssim h \lesssim 12$, more dynamical processes are possible, but the final Type 1 vertex population is limited because dynamics must start at the array edges, and Type 1 vertices that are created near the array edges ``block'' this from happening.

The four different types of disorder studied all have similar effects on the $n_1$ \textit{vs} $h$ curves. In particular, switching disorder has the same effect as energetic disorder. This points to the fact that dynamics are determined through the spin flip criterion \eqref{switching}, which is affected equally by changes to the energy of configurations (through $\vec{h}_{\mathrm{dip}}$) and changes to the switching fields $h_c^{(i)}$. We expect similar results to hold for other switching criteria, such as Stoner-Wohlfarth switching, since the island magnetization reversal always depends on both coupling to other spins and the intrinsic properties of each island.

On the other hand, even though different disorder types have similar effects, the difference between  weak and strong disorder is striking. As expected, in the weak disorder regime disorder acts as a small perturbation on the perfect system. The two non-trivial field regimes are still observed, and $\Delta h$ is changed little from the perfect system.
In contrast, in the strong disorder regime the $n_1$ vs $h$ curve is strongly altered from the perfect system. $\Delta h$ increases to $4$: at very low fields, spins with unfavourable $\vec{h}_{\mathrm{dip}}$ or small $h_c^{(i)}$ can flip, whereas at very high fields the magnetization cannot completely track the rotating field due to pinning of spins with favourable $\vec{h}_{\mathrm{dip}}$ or large $h_c^{(i)}$. The other significant change is that $n_1$ is reduced in the low field regime and increased in the high field regime. This is a result of dynamics being allowed to start in the bulk. In the low field regime, this blocks the regular invasion process that would otherwise lead to large $n_1$. In the high field regime, this allows Type 1-creating dynamics to continue longer, increasing $n_1$ over the perfect case.

In conclusion, different types of disorder in artificial spin ice have very similar effects on simulated dynamics and can all be compared to an energy scale set by the nearest-neighbor coupling. We show elswhere \cite{Budrikis:2011experiments} that this allows disorder in an experimental system to be measured in terms of an effective switching field disorder.

Z.B. and R.L.S. acknowledge the Australian Research Council for funding. Z.B. acknowledges funding from INFN and the Hackett Foundation.


\begin{thebibliography}{3}%
\makeatletter
\providecommand \@ifxundefined [1]{%
 \@ifx{#1\undefined}
}%
\providecommand \@ifnum [1]{%
 \ifnum #1\expandafter \@firstoftwo
 \else \expandafter \@secondoftwo
 \fi
}%
\providecommand \@ifx [1]{%
 \ifx #1\expandafter \@firstoftwo
 \else \expandafter \@secondoftwo
 \fi
}%
\providecommand \natexlab [1]{#1}%
\providecommand \enquote  [1]{``#1''}%
\providecommand \bibnamefont  [1]{#1}%
\providecommand \bibfnamefont [1]{#1}%
\providecommand \citenamefont [1]{#1}%
\providecommand \href@noop [0]{\@secondoftwo}%
\providecommand \href [0]{\begingroup \@sanitize@url \@href}%
\providecommand \@href[1]{\@@startlink{#1}\@@href}%
\providecommand \@@href[1]{\endgroup#1\@@endlink}%
\providecommand \@sanitize@url [0]{\catcode `\\12\catcode `\$12\catcode
  `\&12\catcode `\#12\catcode `\^12\catcode `\_12\catcode `\%12\relax}%
\providecommand \@@startlink[1]{}%
\providecommand \@@endlink[0]{}%
\providecommand \url  [0]{\begingroup\@sanitize@url \@url }%
\providecommand \@url [1]{\endgroup\@href {#1}{\urlprefix }}%
\providecommand \urlprefix  [0]{URL }%
\providecommand \Eprint [0]{\href }%
\providecommand \doibase [0]{http://dx.doi.org/}%
\providecommand \selectlanguage [0]{\@gobble}%
\providecommand \bibinfo  [0]{\@secondoftwo}%
\providecommand \bibfield  [0]{\@secondoftwo}%
\providecommand \translation [1]{[#1]}%
\providecommand \BibitemOpen [0]{}%
\providecommand \bibitemStop [0]{}%
\providecommand \bibitemNoStop [0]{.\EOS\space}%
\providecommand \EOS [0]{\spacefactor3000\relax}%
\providecommand \BibitemShut  [1]{\csname bibitem#1\endcsname}%
\let\auto@bib@innerbib\@empty
\bibitem{Wang:2006} R. F. Wang, \textit{et al.},
Nature
\textbf{439}, 303--306 (2006).%
\bibitem{Qi:2008}%
  \BibitemOpen
  \bibfield  {author} {\bibinfo {author} {\bibfnamefont {Y.}~\bibnamefont
  {Qi}}, \bibinfo {author} {\bibfnamefont {T.}~\bibnamefont {Brintlinger}}, \
  and\ \bibinfo {author} {\bibfnamefont {J.}~\bibnamefont {Cumings}},\
  }
  {\bibfield  {journal} {\bibinfo  {journal}
  {Phys. Rev. B}\ }\textbf
  {\bibinfo {volume} {77}},\ \bibinfo {pages} {94418} (\bibinfo {year}
  {2008})}\BibitemShut {NoStop}%
\bibitem [{\citenamefont {Harris}\ \emph {et~al.}(1997)\citenamefont {Harris},
  \citenamefont {Bramwell}, \citenamefont {McMorrow}, \citenamefont {Zeiske},\
  and\ \citenamefont {Godfrey}}]{1997Harris}%
  \BibitemOpen
  \bibfield  {author} {\bibinfo {author} {\bibfnamefont {M.~J.}\ \bibnamefont
  {Harris}}, \bibinfo {author} {\bibfnamefont {S.~T.}\ \bibnamefont
  {Bramwell}}, \bibinfo {author} {\bibfnamefont {D.~F.}\ \bibnamefont
  {McMorrow}}, \bibinfo {author} {\bibfnamefont {T.}~\bibnamefont {Zeiske}}, \
  and\ \bibinfo {author} {\bibfnamefont {K.~W.}\ \bibnamefont {Godfrey}},\
  }\href@noop {} {\bibfield  {journal} {\bibinfo  {journal} {Phys. Rev. Lett.}\
  }\textbf {\bibinfo {volume} {79}},\ \bibinfo {pages} {2554} (\bibinfo {year}
  {1997})}\BibitemShut {NoStop}%
\bibitem{Libal:2009}A. Lib{\'a}l, C. J. Olson Reichhardt, and C. Reichhardt, Phys. Rev. Lett. \textbf{102}, 237004 (2009).
\bibitem{Budrikis:2011networks}
Z. Budrikis, P. Politi and R. L. Stamps, arXiv:1108.0536 (unpublished).
\bibitem{Kohli:2011}K. K. Kohli, \textit{et al.},
arXiv:1106.1394 (unpublished).
\bibitem{Mellado:2010}P. Mellado, O. Petrova, Y. Shen, and O. Tchernyshyov, Phys. Rev. Lett. \textbf{105}, 187206 (2010).
\bibitem{Ladak:2010}S. Ladak, \textit{et al.}, Nature Phys. \textbf{6}, 359--363 (2010).
\bibitem{Mengotti:2010}E. Mengotti, \textit{et al.}, Nature Phys. \textbf{5}, 68--74 (2010).
\bibitem{Budrikis:2010}Z. Budrikis, P. Politi, and R. L. Stamps,
Phys. Rev. Lett.
\textbf{105}, 017201 (2010).
\bibitem{Budrikis:2011experiments}Z. Budrikis, J. P. Morgan, J. Akerman, A. Stein, R. L. Stamps, P. Politi, S. Langridge, C. H. Marrows, manuscript in preparation.
\end{thebibliography}
\end{document}